\documentclass[reprint,aps,prb,groupedaddress]{revtex4-2}

\usepackage{amsmath,amssymb,amsfonts} 
\usepackage{graphicx}
\usepackage{dcolumn}
\usepackage{bm}
\usepackage[colorlinks=true,citecolor=myblue,linkcolor=myblue,filecolor=myblue,urlcolor=myblue]{hyperref}
\usepackage{braket} 
\usepackage{tikz} 
\usepackage{comment} 
\usepackage[normalem]{ulem} 
\usepackage{booktabs}

\DeclareMathOperator{\Tr}{Tr}

\definecolor{mygreen}{RGB}{0,220,0}
\definecolor{myblue}{RGB}{0, 170, 212}

\begin{document}

\title{Sample complexity of matrix product states at finite temperature}
\author{Atsushi Iwaki}
\email{iwaki-atsushi413@g.ecc.u-tokyo.ac.jp}
\author{Chisa Hotta}
\affiliation{Department of Basic Science, University of Tokyo, Meguro-ku, Tokyo 153-8902, Japan}
\date{\today}

\begin{abstract}
For quantum many-body systems in one dimension, 
computational complexity theory reveals that the evaluation of ground-state energy 
remains elusive on quantum computers, 
contrasting the existence of a classical algorithm for temperatures 
higher than the inverse logarithm of the system size. 
This highlights a qualitative difference 
between low- and high-temperature states in terms of computational complexity. 
Here, we describe finite-temperature states using the matrix product state formalism. 
Within the framework of random samplings, 
we derive an analytical formula for the required number of samples, 
which provides both quantitative and qualitative measures of computational complexity. 
At high and low temperatures, its scaling behavior with system size is linear and quadratic, respectively, 
demonstrating a distinct crossover between these numerically difficult regimes 
of quantitative difference.
\end{abstract}

\maketitle

\section{Introduction}
\label{sec:intro}
The realization of quantum many-body states at finite temperature 
in classical or quantum computers is a pivotal topic 
as it pertains to the modern research areas of 
thermalization \cite{deutsch2018, mori2018}, 
many-body localization \cite{nandkishore2015, abanin2019}, 
quantum many-body scars \cite{moudgalya2022, chandran2023}, 
and so on 
whose clarification often relies heavily on numerical tools. 
Recently, 
quantum simulators have proved to be powerful platforms 
for realizing such states in a large system size in laboratories 
\cite{aidelsburger2013, miyake2013, bernien2017, zhang2017, king2018, harris2018}, 
allowing for comparisons with algorithmic approaches in numerical simulations. 
\par
Computational complexity of physical states 
is related to the degree of difficulty of computational problems 
in relation to the complexity classes, and serves as a guide to such problems. 
One important class is Quantum Merlin-Arthur (QMA) 
which tests whether the polynomial size of a quantum state can be verified within a required polynomial time
in a quantum computer. 
For one-dimensional (1D) quantum many-body states, it is known that 
the evaluation of exact ground-state energy is QMA-complete, indicating that the problem is difficult 
even for quantum computers \cite{kitaev2002, oliveira2008, aharonov2009, nagaj2008, hallgren2013}. 
However, empirically, we can efficiently calculate their ground states very accurately 
for large system sizes even in classical computers 
using e.g. the density matrix renormalization group method \cite{white1992, white1993, schollwock2011}. 
To be precise, such calculations are more established for gapped systems 
\cite{landau2015, chubb2016, arad2017, roberts2017} 
where the entanglement area law holds \cite{hastings2007, eisert2010}, 
and the degrees of difficulty of calculation are closely related to 
whether the system is gapped or gapless. 
However, knowing whether a given Hamiltonian has a gap or not is already an undecidable problem in general 
\cite{cubitt2015, bausch2020}. 
\par
\begin{figure}
    \centering
    \includegraphics[width=\hsize]{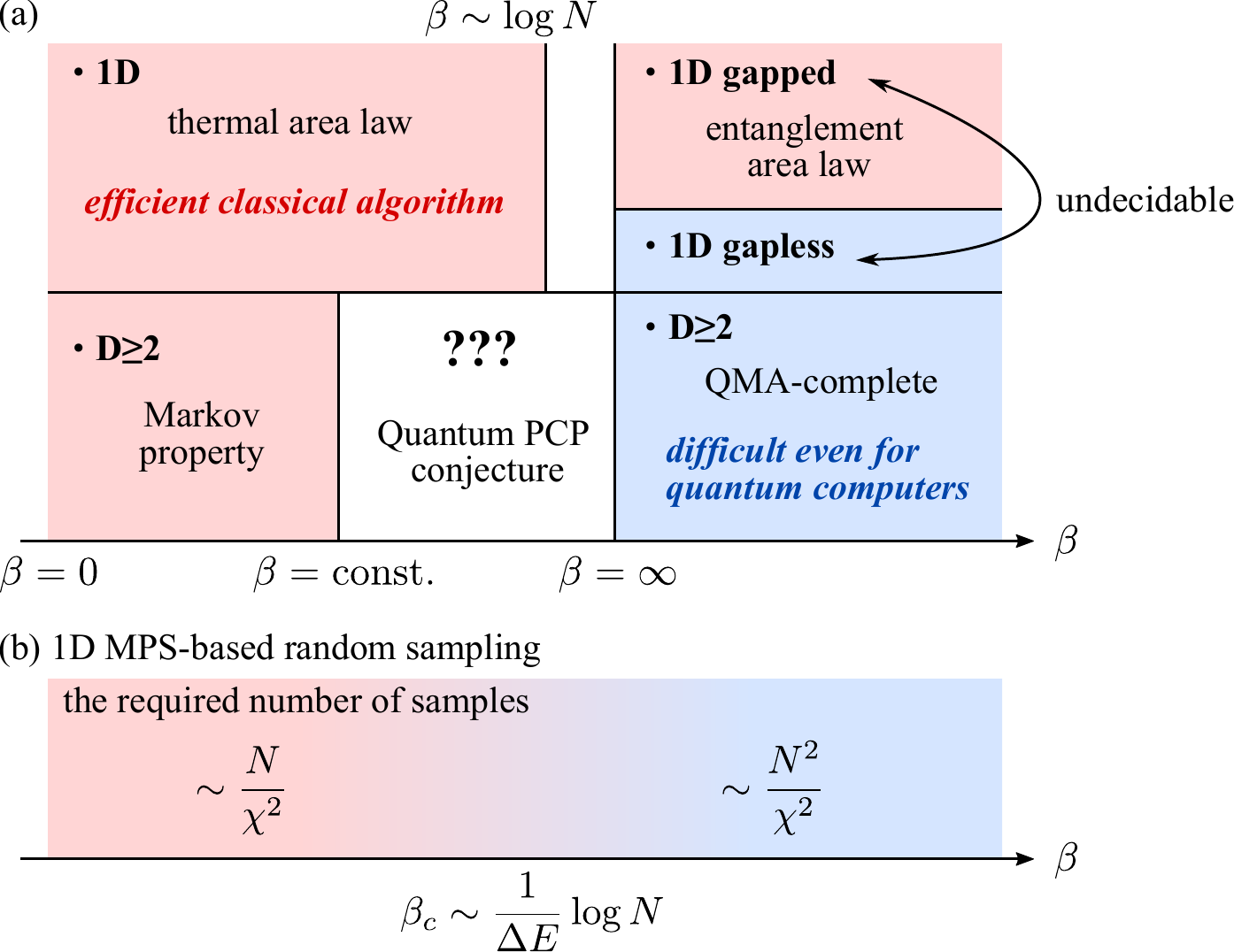}
    \caption{
    (a) Schematic illustration of the computational complexity of thermal and ground states. 
    Regions highlighted in red have efficient classical algorithms, whereas in the blue regions, 
    the ground-state energy cannot be efficiently calculated even for quantum computers in general. 
    (b) Temperature dependence of the required number of samples 
    in MPS-based random sampling methods we derive in this paper. 
    It scales linearly with the system size $N$ at high temperatures, 
    and it is proportional to $N^2$ at low temperatures 
    at leading order, where the initial bond dimension $\chi$ is large.
    The crossover is characterized by the inverse temperature $\beta_c$, 
    where $\Delta E$ is the spectral gap. 
    }
    \label{fig:complexity}
\end{figure}

At finite temperature, preparing thermal states in classical computers is 
relatively easier than preparing the ground state for almost all cases, 
just as it is easy to prepare finite-temperature states in experiments. 
We schematically show in Fig.~\ref{fig:complexity}(a) 
the classification known so far for the computational complexity of states over 
different temperature regimes and different spatial dimensions. 
\begin{table*}
    \centering
    \caption{
    Classification of finite-temperature methods. 
    Physical memory to realize in a computer 
    and the required number of sample averages 
    as a function of the system size $N$ and the bond dimension $\chi$; 
    available $N$ and the applicable spatial dimensions are shown. 
    }
    \begin{ruledtabular}
    \begin{tabular}{lccccc}
        Method & Memory & Sample $\#$ & System size & Spatial dimension & Types of state \\
        \midrule
        Full density operator & $D^2=e^{\mathcal{O}(N)}$ & 1 & $\lesssim 15$ & Any & Gibbs \\
        MPO (density operator) & $Nd\chi^2$ & 1 & $\sim 100$ & 1D or 2D  &  Gibbs \\ 
        \midrule
        TPQ / FTL & $D=e^{\mathcal{O}(N)}$ & $\gtrsim e^{-\mathcal{O}(N)}$ & $\lesssim 30$ & Any & Pure \\
        \midrule
        METTS & $Nd\chi^2$ & $\sim 100?$ & $\sim 100$ & 1D or 2D & Mixed \\
        TPQ-MPS & $Nd\chi^2$ & $\mathcal{O}(N/\chi^2)$ or $\mathcal{O}(N^2/\chi^2)$ & $\sim 100$ & 1D or 2D & Nearly pure \\
    \end{tabular}
    \end{ruledtabular}
    \label{tab:methods}
\end{table*}
Recently, rigorous analysis has progressed in the field of quantum information. 
For 1D systems, the approximate tensor network representation of the Gibbs state 
is obtained when the inverse temperature $\beta$ is smaller than 
the logarithm of the system size $N$, as $\beta \lesssim \log N$ \cite{kuwahara2021, alhambra2021}. 
For higher-dimensional systems, 
it was proved that the partition function can be classically simulated 
in polynomial time with $N$ at temperatures higher than a constant independent of $N$ 
\cite{kuwahara2020, harrow2020}. 
Phase transitions can occur only in higher spatial dimensions at temperatures lower than discussed here, 
but this region is not well understood. 
It is considered to be related to an open problem known as the quantum PCP conjecture
in the context of computational complexity theory \cite{aharonov2013}. 
\par
Practical numerical methods have a long history in condensed matter physics. 
The stochastic quantum Monte Carlo (QMC) method allows large system sizes $N$ 
but faces a sign problem \cite{foulkes2001, sandvik2010}. 
High-temperature expansions \cite{domb1974, oitmaa2006} and 
numerical linked-cluster expansions \cite{rigol2006, rigol2007-1, rigol2007-2} 
have been established, although the former has a temperature bound of $\beta^{-1} \lesssim J$ 
for a typical energy scale $J$ and the latter fails when the correlation length diverges. 
\par
Methods directly describing thermal states have more variants (see Table~\ref{tab:methods}). 
The most conventional Gibbs state 
requires computational memory of $D^2$, where $D$ is the dimension of the Hilbert space 
because it is a maximally mixed state 
that is represented by the full-rank density operator. 
Thermal equilibrium can be represented by a single pure state as well. 
The finite-temperature Lanczos (FTL) methods can handle it using an appropriate choice of basis with 
a computational memory of $D$ \cite{imada1986, jaklic1994, hams2000, schnack2020}, 
and another is called the thermal pure quantum (TPQ) method \cite{sugiura2012, sugiura2013}. 
Because of the concept of typicality \cite{popescu2006, goldstein2006, reimann2007, sugita2007}, 
it is guaranteed that the required number of samples is very small, 
whereas the available $N$ is limited by the growth of $D=d^N$, where $d$ is the local dimension. 
To access large systems, 
tensor network methods have been actively studied. 
Particularly, matrix product states (MPSs) are commonly used in 1D 
\cite{fannes1992}, 
taking advantage of the time-evolving block decimation (TEBD) algorithm \cite{vidal2004, white2004, daley2004}. 
For the Gibbs state, with the doubled Hilbert space, 
the matrix product operator (MPO) \cite{verstraete2004-2, zwolak2004}, 
the purification represented by the MPS \cite{feiguin2005} 
and its analogs \cite{chen2018, hauschild2018, li2023} exist. 
Besides the Gibbs and TPQ states, we have many thermal states in between; 
we previously developed the TPQ-MPS method \cite{iwaki2021, gohlke2023} for the nearly pure state (see Sec.\ref{subsec:tpq-mps}), 
which belongs to the family of MPS-based random sampling methods
\cite{garnerone2013-1, garnerone2013-2, goto2021, gao2023}, 
but provides the purest thermal state. 
The minimally entangled typical thermal states (METTS) method \cite{white2009, stoudenmire2010} 
has less purity than TPQ-MPS, although its purity is being improved by different devices 
\cite{bruognolo2015, binder2015, binder2017, goto2020, chen2020, chung2019}. 
\par
In principle, all these finite-temperature numerical approaches 
start from a high temperature and, in approaching a low temperature, suffer numerical difficulties. 
Many of them belong to the random sampling method, 
which prepares the initial random states at high temperature and cools them down by the imaginary time evolution. 
In that framework, the numerical difficulty can be measured by the required number of samples, 
which we call \textit{``sample complexity"}. 
The sample complexity crucially depends on the expression power of the method for the sampled states. 
Using this idea, we recently classified the Gibbs state, the TPQ state and the states in between them 
as thermal mixed quantum (TMQ) states \cite{iwaki2022}. 
The Gibbs state method that can basically express a (semiclassical) product state requires a large sampling number, 
while the TPQ methods that can express quantum states with large enough entanglement require only a few samples. 
In Table~\ref{tab:methods}, we show the classification of methods in this context; 
the expression power of the quantum state 
is the ability to store entropy in the form of quantum entanglement in a single sample, 
because the thermal entropy (stored as the number of samples) 
and the quantum entanglement entropy are two sides of the same coin. 
\par
Another factor that dictates the numerical difficulty 
of representing the thermal states with MPSs is the bond dimension $\chi$. 
Several studies based on rigorous analysis suggested that 
$\chi$ is bounded by an exponential with $\beta$ 
\cite{hastings2006, kliesch2014, molnar2015, kuwahara2021}, 
whereas it often happens that this bound is practically an overestimation. 
In fact, conformal field theory (CFT) analysis showed that $\chi$ grows polynomially with $\beta$ 
\cite{dubail2017, barthel2017}, 
which was confirmed numerically \cite{znidaric2008, kusuki2023}.
\par
In this paper, we investigate the required number of samples in MPS-based methods, 
focusing on the TPQ-MPS method which utilizes a natural description of a TMQ state very close to the TPQ state. 
There is a convenient and measurable quantity that has an amplitude 
proportional to the required number of samples 
called the normalized fluctuation of partition function (NFPF) \cite{iwaki2022}. 
Here, we propose an analytical formula for the NFPF which applies from zero to the highest temperature. 
Importantly, the leading orders of the NFPF or the numbers of random samples 
in terms of $N$ are different between low- and high-temperature limits 
[see Fig.~\ref{fig:complexity}(b)]. 
With numerical demonstrations for several quantum spin models, 
we verify this formula and demonstrate the explicit crossover 
between the two temperature regions, 
which are qualitatively different in the 
language of computational complexity.

\section{Random sampling methods and NFPF}
The physical quantity in the thermal equilibrium is described as 
\begin{equation}
    \langle O \rangle_\beta 
    = \Tr[\rho_\mathrm{G}(\beta) O], 
\end{equation}
where a density operator $\rho_\mathrm{G}(\beta)$ represents 
the Gibbs state and is given as 
\begin{equation}
    \rho_\mathrm{G}(\beta) = \frac{e^{-\beta H}}{Z(\beta)}, \quad 
    Z(\beta) = \Tr e^{-\beta H}. 
\end{equation}
Here, we introduce the framework of the random sampling method 
for finite-temperature calculations and a way to evaluate  
their efficiency \cite{goto2021} based on Ref.~[\onlinecite{iwaki2022}]. 
This framework applies to Monte Carlo methods without using a Markov chain 
but does not include well-established Markov chain Monte Carlo methods such as QMC and METTS. 
However, its applicability is wide and is not only limited to 
finite-temperature calculations; it also applies to other quantum states whose realization follows 
distributions other than the Boltzmann distribution. 
\subsection{Random sampling methods}
We initially prepare a random state $\ket{\psi_0}$ that satisfies
\begin{equation}
    \overline{\ket{\psi_0}\bra{\psi_0}} = c I, 
\end{equation}
where $\overline{\cdots}$ denotes taking the random average 
and $c$ is a system-size-dependent constant, 
determined naturally 
by the choice of $\ket{\psi_0}$'s, which are kept unnormalized for a reason we will see shortly.
This equation implies that $\ket{\psi_0}$ represents one of the states realized in the high-temperature limit. 
In practice, what temperature it targets depends on the quality of the random sampling and 
the approximations used for the representation of $\ket{\psi_0}$. 
We perform an imaginary time evolution to cool the temperature 
to $\beta^{-1}$ to obtain the corresponding state 
\begin{align}
   \ket{\psi_\beta} = e^{- \beta H/2} \ket{\psi_0}, 
\label{eq:imagev}
\end{align}
where $H$ is the system Hamiltonian. 
The partition function or certain physical quantities are obtained by taking the random average: 
\begin{gather}
   Z(\beta) = \frac{1}{c} \overline{\braket{\psi_\beta|\psi_\beta}}, 
   \label{eq:zz} \\
   \langle O \rangle_\beta = \frac{\overline{\braket{\psi_\beta|O|\psi_\beta}}}{\overline{\braket{\psi_\beta|\psi_\beta}}}. 
\end{gather}
As the ideal random averages are not accessible, 
we approximate them by using sample averages in the numerical simulations as
\begin{gather}
   Z^\mathrm{samp}_M (\beta) = \frac{1}{cM} \sum_{i=1}^M \braket{\psi_\beta^{(i)}|\psi_\beta^{(i)}} ,
\label{eq:zsamp}\\
   \langle O \rangle^\mathrm{samp}_{\beta, M} =
   \frac{\sum_{i=1}^M \braket{\psi_\beta^{(i)}|O|\psi_\beta^{(i)}}}
   {\sum_{j=1}^M \braket{\psi_\beta^{(j)}|\psi_\beta^{(j)}}},
\label{eq:osamp}
\end{gather}
where $\{\ket{\psi_\beta^{(i)}}\}_{i=1}^M$ are $M$ independent realizations of $\ket{\psi_\beta}$. 
If we take a sufficiently large value of $M$, 
the law of large numbers guarantees that the sample average matches the random average. 
In Eq.~\eqref{eq:osamp}, the norm $\braket{\psi_\beta|\psi_\beta}$ of each sample 
serves as the weight in the sample average, 
and the larger weight means that the state after the time evolution remains closer to the ideal thermal state. 
When, for example, expanded as $\ket{\psi_0}=\sum_n a_n \ket{n}$ with a proper basis set $\{\ket{n}\}$, 
the raw values of random coefficients $\{ a_n \}$ belonging to different samples include information 
about their relative importance. 
For this reason, the initial norm $\braket{\psi_0|\psi_0}$ has a physical meaning, 
and $\ket{\psi_0}$ should be kept un-normalized
\footnote{In the first proposal of TPQ-MPS in Ref.~[\onlinecite{iwaki2021}], we chose to normalize $\ket{\psi_0}$, 
which was updated to the un-normalized version to give better quality of random sampling methods.}.
\par
Conventionally, the sufficient sample number $M$ required was empirically determined depending 
on the physical quantities one wanted to obtain to sufficiently reduce the variance 
according to the objective. 
However, this ambiguity can be removed by using a quantity we recently introduced, 
the NFPF \cite{iwaki2022}. 
By using the NFPF, we can decisively discuss $M$ as not only the sufficient but also the necessary value 
to qualify the thermal quantum state on equal footing across different numerical methods we apply 
in operating the random sampling method. 
The value of $M$ or the NFPF depends on how the initial random states are prepared 
and on the approximations of the method. 
\par
The NFPF is given as 
\begin{align}
   \delta z^2 = \frac{\mathrm{Var}\left(\braket{\psi_\beta|\psi_\beta}\right)}{\left(\overline{\braket{\psi_\beta|\psi_\beta}}\right)^2}. 
\end{align}
It quantifies the random fluctuation of the partition function, $Z(\beta) = \Tr e^{-\beta H}$, as 
\begin{align}
   \frac{1}{Z(\beta)^2} \overline{\left[Z^\mathrm{samp}_M (\beta) - Z(\beta)\right]^2}
   = \frac{\delta z^2}{M}. 
\end{align}
Therefore, to obtain the partition function with a relative error $\epsilon$, 
we require $M_\epsilon$ samples, given by
\begin{align}
   M_\epsilon = \frac{\delta z^2}{\epsilon ^2}.
\end{align}
With Chebyshev's inequality, we can ascribe a strictly probabilistic interpretation to the value of $M_\epsilon$: 
\begin{align}
   \mathrm{Prob}\left[|Z^\mathrm{samp}_M (\beta) - Z(\beta)|
   \ge \epsilon Z(\beta) \right] \le \frac{M_\epsilon}{M}. 
\end{align}
Under the condition 
\begin{align}
   \mathrm{Var}\left(\braket{\psi_\beta|O|\psi_\beta}\right) \le \mathrm{(const.)} \times 
   \|O\|^2 \mathrm{Var}(\braket{\psi_\beta|\psi_\beta}),  
   \label{eq:condition}
\end{align}
the NFPF gives a bound of the fluctuation of the physical quantity $O$ as 
\begin{align}
   \overline{\left(\langle O \rangle^\mathrm{samp}_{\beta, M} 
   - \langle O \rangle_\beta\right)^2} 
   \lesssim \mathrm{(const.)} \times \| O \|^2 \frac{\delta z^2}{M}. 
\end{align}
This condition typically holds except in a few specific cases, 
and indeed the required sample size for the evaluation of energy is proportional to $M_\epsilon$ \cite{iwaki2022}. 

\subsection{NFPF of the TPQ state}
As an ideal reference system, 
we consider the TPQ method, which uses the entire Hilbert space with the dimension $D$ to describe the thermal state. 
There are several different ways to prepare initial random states in the TPQ method \cite{jin2021}. 
We choose to generate the initial state to have the coefficients of the basis 
following an independent complex Gaussian distribution. 
Importantly, this random state is independent of the choice of the basis and thus can be expressed as
$\ket{\psi_0} = \sum_n a_n \ket{n}$, 
where $\{\ket{n}\}_n$ is the energy eigenstates. 
With this setting, we can calculate the NFPF of the TPQ method analytically as
\begin{align}
   \delta z^2_\mathrm{TPQ} = e ^{-Ns_2 (\beta)}, 
\end{align}
where $s_2(\beta)$ is the rescaled R\'{e}nyi-2 entropy of the Gibbs state, 
\begin{align}
   s_2(\beta) = - \frac{1}{N} \log \Tr[\rho_\mathrm{G}(\beta)^2],  
\end{align}
which is called the thermal R\'{e}nyi-2 entropy. 
Since the NFPF decreases exponentially with the system size, 
the TPQ method proves to be highly efficient from the perspective of sample complexity. 
Moreover, inequality~\eqref{eq:condition} holds for all physical quantities in the TPQ state. 
As a consequence, random fluctuations of physical quantities in the TPQ method are significantly reduced, 
and for appropriately large systems, only a few samples are necessary.

\subsection{TPQ-MPS method} 
\label{subsec:tpq-mps}
We briefly explain the TPQ-MPS method \cite{iwaki2021}. 
Unlike the standard MPS form with open edges where the interactions are absent, 
we apply the specific form of the MPS that hosts two auxiliary sites located on both edges. 
The advantage of this construction was already proved in two previous works; 
in Ref.~[\onlinecite{iwaki2021}], we proposed the TPQ-MPS method and showed the volume law entanglement in 1D, 
and in Ref.~[\onlinecite{gohlke2023}], we successfully applied it to the two-dimensional (2D) honeycomb Kitaev model.
In 1D \cite{iwaki2021}, the number of sample averages $M$ decreases significantly with decreasing the temperature, 
and the variance of measured quantities is suppressed at lower temperatures, 
in contrast to all the other known methods, including QMC and the standard TPQ method 
(or the FTL method) 
using the full Hilbert space. 
The reason is as follows; 
in typical MPSs, the bond dimension is 1 at both edges and increases as $d^n$ as we approach 
$n=1,2,\cdots$ toward the center site, 
which is known to be reflected in the Page curve of the entanglement entropy. 
This means that due to the boundary condition, the amount of entanglement entropy stored in the system is strongly 
suppressed toward the edge sites. 
However, in our construction, auxiliaries helps us to avoid such suppression, and the bond dimension 
and the entanglement entropy stay almost constant regardless of the location of the bipartition of the system. 
The amount of entropy stored thus follows the volume law that accounts for that of the actual thermal entropy 
of the system, allowing the system to be nearly pure. 
\par
The initial state is represented as a random MPS with auxiliaries as 
\begin{align}
   \ket{\psi_0}
   = \sum_{\{\alpha\}}^\chi \sum_{\{i\}}^d
   & A_{\alpha_0 \alpha_1}^{[1]i_1} A_{\alpha_1 \alpha_2}^{[2]i_2} \cdots
   A_{\alpha_{N-1} \alpha_N}^{[N]i_N} \notag \\
   & \times \ket{\alpha_0, i_1, i_2, \dots , i_N, \alpha_N}. 
\label{eq:tpqmps}
\end{align}
The auxiliary systems have the same degrees of freedom as the initial bond dimension $\chi$. 
We then apply an imaginary time evolution, Eq.~\eqref{eq:imagev}, and obtain $\ket{\psi_\beta}$ 
for this construction and evaluate the physical quantities 
following Eqs.~\eqref{eq:zsamp} and \eqref{eq:osamp}. 
The bond dimension may increase through imaginary time evolution 
using the TEBD technique or by applying the MPO. 
Previously, we showed that the information on higher-temperature states is properly discarded 
through the truncation process in the TPQ-MPS method \cite{gohlke2023}. 

\section{Sample complexity of matrix product states}
In this section, we analytically derive a formula for the NFPF of the TPQ-MPS. 
By using the formula for the ideal random average distribution, 
we are able to perform the analytical evaluation of the related quantities. 
\par
To simplify the formula, we utilize diagram notations of tensor networks. 
We first focus on Eq.~\eqref{eq:tpqmps} 
depicted as 
\begin{align}
    \ket{\psi_0}
    &= 
    \raisebox{-0.5\height}{
      \begin{tikzpicture}[x=3mm,y=3mm]
         \draw[line width=1.5pt,rounded corners=7pt]
            (0,-2)--(0,0)--(7,0);
         \draw[line width=1.5pt,rounded corners=7pt]
            (9,0)--(13,0)--(13,-2);
         \foreach \x in {7.5, 8, 8.5}{
            \draw[fill=black] (\x, 0) circle [radius=0.05];
         }
         \draw (2,0)--(2,-2);
         \draw (5,0)--(5,-2);
         \draw (11,0)--(11,-2);
         \draw[fill=red!30] (2,0) circle (1);
         \draw[fill=red!50!yellow!30] (5,0) circle (1);
         \draw[fill=yellow!30] (11,0) circle (1);
      \end{tikzpicture} 
   } 
\end{align}
where the circle represents a matrix $A$ 
and thin (thick) lines mean physical (auxiliary) degrees of freedom. 
We colored the matrices (circles) differently implying that each matrix follows 
an independent probability distribution. 
We assume that all elements of matrices obey an independent complex Gaussian distribution. 
A tensor $a_i$ which is taken from an independent complex Gaussian distribution satisfies
\begin{equation}
   \overline{a_i a_j^*} = \delta_{ij}, \quad
   \overline{a_i a_j^* a_k a_l^*} 
   = \delta_{ij} \delta_{kl} + \delta_{il} \delta_{jk}. 
   \label{eqs:Gauss_dist}
\end{equation}
These equations are represented in the following tensor network diagrams. 
\begin{align}
   \overline{\left[\raisebox{-0.45\height}{
      \begin{tikzpicture}[x=2mm,y=2mm]
         \draw (1,0)--(1,2);
         \draw (1,5)--(1,7);
         \draw[fill=red!20] (1,2) circle (1.2) node {$a^*$};
         \draw[fill=red!20] (1,5) circle (1.2) node {$a$};
      \end{tikzpicture}
   }\right]}
   =
   \raisebox{-0.5\height}{
      \begin{tikzpicture}[x=2mm,y=2mm]
         \draw (1,0)--(1,7);
      \end{tikzpicture}
   }
   , \quad 
   \overline{\left[\raisebox{-0.45\height}{
      \begin{tikzpicture}[x=2mm,y=2mm]
         \draw (1,0)--(1,2);
         \draw (1,5)--(1,7);
         \draw[fill=red!20] (1,2) circle (1.2) node {$a^*$};
         \draw[fill=red!20] (1,5) circle (1.2) node {$a$};
      \end{tikzpicture}
      \begin{tikzpicture}[x=2mm,y=2mm]
         \draw (1,0)--(1,2);
         \draw (1,5)--(1,7);
         \draw[fill=red!20] (1,2) circle (1.2) node {$a^*$};
         \draw[fill=red!20] (1,5) circle (1.2) node {$a$};
      \end{tikzpicture}
   }\right]}
   =
   \raisebox{-0.45\height}{
      \begin{tikzpicture}[x=2mm,y=2mm]
         \draw (0,0)--(0,7);
         \draw (3,0)--(3,7);
      \end{tikzpicture}
   }
   +
   \raisebox{-0.45\height}{
      \begin{tikzpicture}[x=2mm,y=2mm]
         \draw (0,0)--(3,7);
         \draw (3,0)--(0,7);
      \end{tikzpicture}
   }.
   \label{eq:gauss_rule}
\end{align}
We will calculate the NFPF of the TPQ-MPS using the above relations. 

\subsection{Non-interacting Hamiltonian}
\label{subsec:non-interacting}
Here, we consider a Hamiltonian which is the sum of single-site local operators with translational invariance, 
\begin{align}
    H = \sum_{i=1}^N h_i, \quad
   h_i = \overset{1}{I} \otimes \cdots \otimes \overset{i}{h} 
   \otimes \cdots \otimes \overset{N}{I}.
\end{align}
The partition function is defined at each local site independently of the other sites as 
$Z(\beta) = z(\beta)^N$, with $z(\beta) = \Tr e^{-\beta h}$.
To evaluate the NFPF, we initially calculate the norm of the finite-temperature state, 
which is expressed as 
\begin{align}
   \braket{\psi_\beta|\psi_\beta} = 
   \raisebox{-0.45\height}{
      \begin{tikzpicture}[x=3mm,y=3mm]
         \draw[line width=1.6pt,rounded corners=7pt]
            (7,1)--(0,1)--(0,7)--(7,7);
         \draw[line width=1.6pt,rounded corners=7pt]
            (9,7)--(13,7)--(13,1)--(9,1);
         \foreach \x in {7.5, 8, 8.5}{
            \draw[fill=black] (\x, 7) circle [radius=0.05];
         }
         \foreach \x in {7.5, 8, 8.5}{
            \draw[fill=black] (\x, 1) circle [radius=0.05];
         }
         \draw (2,1)--(2,7);
         \draw (5,1)--(5,7);
         \draw (11,1)--(11,7);
         \draw[fill=red!30] (2,7) circle (1);
         \draw[fill=red!50!yellow!30] (5,7) circle (1);
         \draw[fill=yellow!30] (11,7) circle (1);
         \draw[fill=red!30] (2,1) circle (1) node {$*$};
         \draw[fill=red!50!yellow!30] (5,1) circle (1) node {$*$};
         \draw[fill=yellow!30] (11,1) circle (1) node {$*$};
         \draw[fill=blue!20] (1,3) rectangle (3,5);
         \draw[fill=blue!20] (4,3) rectangle (6,5);
         \draw[fill=blue!20] (10,3) rectangle (12,5);
      \end{tikzpicture}
   },
\end{align}
where $ 
\raisebox{-0.3\height}{
   \begin{tikzpicture}[x=1.5mm,y=1.5mm]
      \draw (1,0)--(1,4);
      \draw[fill=blue!20] (0,1) rectangle (2,3);
   \end{tikzpicture}
}
$ represents the operator $e^{-\beta h}$. 
By applying Eq.~\eqref{eq:gauss_rule} to each matrix, 
we can calculate the random average of the norm as 
\begin{align}
   \overline{\braket{\psi_\beta|\psi_\beta}} 
   &= 
   \raisebox{-0.45\height}{
      \begin{tikzpicture}[x=3mm,y=3mm]
         \draw[line width=1.6pt,rounded corners=3mm]
            (1,0)--(1,4)--(3,4)--(3,0)--cycle;
      \end{tikzpicture}
      \begin{tikzpicture}[x=3mm,y=3mm]
         \draw[rounded corners=3mm]
            (1,0)--(1,4)--(3,4)--(3,0)--cycle;
         \draw[fill=blue!20] (0,1) rectangle (2,3);
      \end{tikzpicture}
      \begin{tikzpicture}[x=3mm,y=3mm]
         \draw[line width=1.6pt,rounded corners=3mm]
            (1,0)--(1,4)--(3,4)--(3,0)--cycle;
      \end{tikzpicture}
      \begin{tikzpicture}[x=3mm,y=3mm]
         \draw[rounded corners=3mm]
            (1,0)--(1,4)--(3,4)--(3,0)--cycle;
         \draw[fill=blue!20] (0,1) rectangle (2,3);
      \end{tikzpicture}
   }
   \cdots
   \raisebox{-0.45\height}{
      \begin{tikzpicture}[x=3mm,y=3mm]
         \draw[rounded corners=3mm]
            (1,0)--(1,4)--(3,4)--(3,0)--cycle;
         \draw[fill=blue!20] (0,1) rectangle (2,3);
      \end{tikzpicture}
      \begin{tikzpicture}[x=3mm,y=3mm]
         \draw[line width=1.6pt,rounded corners=3mm]
            (1,0)--(1,4)--(3,4)--(3,0)--cycle;
      \end{tikzpicture}
   } \notag \\
   &= \chi^{N+1} z(\beta)^N = \chi^{N+1} Z(\beta).
\end{align}
This calculation verifies Eq.~\eqref{eq:zz}. 
In the TPQ-MPS method, it is found that $c=\chi^{N+1}$.
Next, we calculate the random average of the square of the norm: 
\begin{align}
   &\overline{\braket{\psi_\beta|\psi_\beta}^2} \notag \\ 
   &= 
   \overline{\left[\raisebox{-0.47\height}{
      \begin{tikzpicture}[x=2mm,y=2mm]
         \draw[line width=1.6pt,rounded corners=7pt]
            (7,1)--(0,1)--(0,7)--(7,7);
         \draw[line width=1.6pt,rounded corners=7pt]
            (9,7)--(13,7)--(13,1)--(9,1);
         \foreach \x in {7.5, 8, 8.5}{
            \draw[fill=black] (\x, 7) circle [radius=0.05];
         }
         \foreach \x in {7.5, 8, 8.5}{
            \draw[fill=black] (\x, 1) circle [radius=0.05];
         }
         \draw (2,1)--(2,7);
         \draw (5,1)--(5,7);
         \draw (11,1)--(11,7);
         \draw[line width=1.6pt,rounded corners=7pt]
            (7,1+9)--(0,1+9)--(0,7+9)--(7,7+9);
         \draw[line width=1.6pt,rounded corners=7pt]
            (9,7+9)--(13,7+9)--(13,1+9)--(9,1+9);
         \foreach \x in {7.5, 8, 8.5}{
            \draw[fill=black] (\x, 7+9) circle [radius=0.05];
         }
         \foreach \x in {7.5, 8, 8.5}{
            \draw[fill=black] (\x, 1+9) circle [radius=0.05];
         }
         \draw (2,1+9)--(2,7+9);
         \draw (5,1+9)--(5,7+9);
         \draw (11,1+9)--(11,7+9);
         \draw[fill=red!30] (2,7) circle (1);
         \draw[fill=red!50!yellow!30] (5,7) circle (1);
         \draw[fill=yellow!30] (11,7) circle (1);
         \draw[fill=red!30] (2,1) circle (1) node {$*$};
         \draw[fill=red!50!yellow!30] (5,1) circle (1) node {$*$};
         \draw[fill=yellow!30] (11,1) circle (1) node {$*$};
         \draw[fill=red!30] (2, 7+9) circle (1);
         \draw[fill=red!50!yellow!30] (5, 7+9) circle (1);
         \draw[fill=yellow!30] (11,7+9) circle (1);
         \draw[fill=red!30] (2, 1+9) circle (1) node {$*$};
         \draw[fill=red!50!yellow!30] (5, 1+9) circle (1) node {$*$};
         \draw[fill=yellow!30] (11,1+9) circle (1) node {$*$};
         \draw[fill=blue!20] (1, 3) rectangle (3, 5);
         \draw[fill=blue!20] (4, 3) rectangle (6, 5);
         \draw[fill=blue!20] (10,3) rectangle (12,5);
         \draw[fill=blue!20] (1, 3+9) rectangle (3, 5+9);
         \draw[fill=blue!20] (4, 3+9) rectangle (6, 5+9);
         \draw[fill=blue!20] (10,3+9) rectangle (12,5+9);
      \end{tikzpicture}
   }\right]} \notag \\
   &= 
   \raisebox{-0.45\height}{
      \begin{tikzpicture}[x=2mm,y=2mm]
         \draw[line width=1.6pt,rounded corners=5pt]
            (1,0)--(0,0)--(0,4)--(1,4);
         \draw[line width=1.6pt,rounded corners=5pt]
            (1,0+5)--(0,0+5)--(0,4+5)--(1,4+5);
      \end{tikzpicture}
   }
   \left[
   \raisebox{-0.45\height}{
      \begin{tikzpicture}[x=2mm,y=2mm]
         \draw[line width=1.6pt,rounded corners=5pt]
            (0,0)--(1,0)--(1,4)--(0,4);
         \draw[line width=1.6pt,rounded corners=5pt]
            (0,0+5)--(1,0+5)--(1,4+5)--(0,4+5);
      \end{tikzpicture}
      \begin{tikzpicture}[x=2mm,y=2mm]
         \draw[rounded corners=5pt]
            (3,0)--(3,4)--(5,4)--(5,0)--cycle;
         \draw[fill=blue!20] (2,1) rectangle (4,3);
         \draw[rounded corners=5pt]
            (3,0+5)--(3,4+5)--(5,4+5)--(5,0+5)--cycle;
         \draw[fill=blue!20] (2,1+5) rectangle (4,3+5);
      \end{tikzpicture}
      \begin{tikzpicture}[x=2mm,y=2mm]
         \draw[line width=1.6pt,rounded corners=5pt]
            (7,0)--(6,0)--(6,4)--(7,4);
         \draw[line width=1.6pt,rounded corners=5pt]
            (7,0+5)--(6,0+5)--(6,4+5)--(7,4+5);
      \end{tikzpicture}
   }
   +
   \raisebox{-0.45\height}{
      \begin{tikzpicture}[x=2mm,y=2mm]
         \draw[line width=1.6pt,rounded corners=5pt]
            (0,0)--(1,0)--(1,9)--(0,9);
         \draw[line width=1.6pt,rounded corners=1pt]
            (0,4)--(0.5,4)--(0.5, 5)--(0,5);
      \end{tikzpicture}
      \begin{tikzpicture}[x=2mm,y=2mm]
         \draw[rounded corners=5pt]
            (3,0)--(3,9)--(5,9)--(5,0)--cycle;
         \draw[fill=blue!20] (2,1) rectangle (4,3);
         \draw[fill=blue!20] (2,1+5) rectangle (4,3+5);
      \end{tikzpicture}
      \begin{tikzpicture}[x=2mm,y=2mm]
         \draw[line width=1.6pt,rounded corners=5pt]
            (1,0)--(0,0)--(0,9)--(1,9);
         \draw[line width=1.6pt,rounded corners=1pt]
            (1,4)--(0.5,4)--(0.5, 5)--(1,5);
      \end{tikzpicture}
   }
   \right]
   \times \cdots \notag \\
   & \times 
   \left[
   \raisebox{-0.45\height}{
      \begin{tikzpicture}[x=2mm,y=2mm]
         \draw[line width=1.6pt,rounded corners=5pt]
            (0,0)--(1,0)--(1,4)--(0,4);
         \draw[line width=1.6pt,rounded corners=5pt]
            (0,0+5)--(1,0+5)--(1,4+5)--(0,4+5);
      \end{tikzpicture}
      \begin{tikzpicture}[x=2mm,y=2mm]
         \draw[rounded corners=5pt]
            (3,0)--(3,4)--(5,4)--(5,0)--cycle;
         \draw[fill=blue!20] (2,1) rectangle (4,3);
         \draw[rounded corners=5pt]
            (3,0+5)--(3,4+5)--(5,4+5)--(5,0+5)--cycle;
         \draw[fill=blue!20] (2,1+5) rectangle (4,3+5);
      \end{tikzpicture}
      \begin{tikzpicture}[x=2mm,y=2mm]
         \draw[line width=1.6pt,rounded corners=5pt]
            (7,0)--(6,0)--(6,4)--(7,4);
         \draw[line width=1.6pt,rounded corners=5pt]
            (7,0+5)--(6,0+5)--(6,4+5)--(7,4+5);
      \end{tikzpicture}
   }
   +
   \raisebox{-0.45\height}{
      \begin{tikzpicture}[x=2mm,y=2mm]
         \draw[line width=1.6pt,rounded corners=5pt]
            (0,0)--(1,0)--(1,9)--(0,9);
         \draw[line width=1.6pt,rounded corners=1pt]
            (0,4)--(0.5,4)--(0.5, 5)--(0,5);
      \end{tikzpicture}
      \begin{tikzpicture}[x=2mm,y=2mm]
         \draw[rounded corners=5pt]
            (3,0)--(3,9)--(5,9)--(5,0)--cycle;
         \draw[fill=blue!20] (2,1) rectangle (4,3);
         \draw[fill=blue!20] (2,1+5) rectangle (4,3+5);
      \end{tikzpicture}
      \begin{tikzpicture}[x=2mm,y=2mm]
         \draw[line width=1.6pt,rounded corners=5pt]
            (1,0)--(0,0)--(0,9)--(1,9);
         \draw[line width=1.6pt,rounded corners=1pt]
            (1,4)--(0.5,4)--(0.5, 5)--(1,5);
      \end{tikzpicture}
   }
   \right]
   \raisebox{-0.45\height}{
      \begin{tikzpicture}[x=2mm,y=2mm]
         \draw[line width=1.6pt,rounded corners=5pt]
            (0,0)--(1,0)--(1,4)--(0,4);
         \draw[line width=1.6pt,rounded corners=5pt]
            (0,0+5)--(1,0+5)--(1,4+5)--(0,4+5);
      \end{tikzpicture}
   }.
\end{align}
This calculation can be recognized within Temperley-Lieb algebra \cite{temperley1971}. 
For simplicity, we define two symbols, $\mathcal{A}$ and $\mathcal{B}$, as
\begin{equation}
    \mathcal{A} =
    \raisebox{-0.45\height}{
      \begin{tikzpicture}[x=2mm,y=2mm]
         \draw[line width=1.6pt,rounded corners=5pt]
            (0,0)--(1,0)--(1,4)--(0,4);
         \draw[line width=1.6pt,rounded corners=5pt]
            (0,0+5)--(1,0+5)--(1,4+5)--(0,4+5);
      \end{tikzpicture}
      \begin{tikzpicture}[x=2mm,y=2mm]
         \draw[rounded corners=5pt]
            (3,0)--(3,4)--(5,4)--(5,0)--cycle;
         \draw[fill=blue!20] (2,1) rectangle (4,3);
         \draw[rounded corners=5pt]
            (3,0+5)--(3,4+5)--(5,4+5)--(5,0+5)--cycle;
         \draw[fill=blue!20] (2,1+5) rectangle (4,3+5);
      \end{tikzpicture}
      \begin{tikzpicture}[x=2mm,y=2mm]
         \draw[line width=1.6pt,rounded corners=5pt]
            (7,0)--(6,0)--(6,4)--(7,4);
         \draw[line width=1.6pt,rounded corners=5pt]
            (7,0+5)--(6,0+5)--(6,4+5)--(7,4+5);
      \end{tikzpicture}
   } , \quad
   \mathcal{B} = 
   \raisebox{-0.45\height}{
      \begin{tikzpicture}[x=2mm,y=2mm]
         \draw[line width=1.6pt,rounded corners=5pt]
            (0,0)--(1,0)--(1,9)--(0,9);
         \draw[line width=1.6pt,rounded corners=1pt]
            (0,4)--(0.5,4)--(0.5, 5)--(0,5);
      \end{tikzpicture}
      \begin{tikzpicture}[x=2mm,y=2mm]
         \draw[rounded corners=5pt]
            (3,0)--(3,9)--(5,9)--(5,0)--cycle;
         \draw[fill=blue!20] (2,1) rectangle (4,3);
         \draw[fill=blue!20] (2,1+5) rectangle (4,3+5);
      \end{tikzpicture}
      \begin{tikzpicture}[x=2mm,y=2mm]
         \draw[line width=1.6pt,rounded corners=5pt]
            (1,0)--(0,0)--(0,9)--(1,9);
         \draw[line width=1.6pt,rounded corners=1pt]
            (1,4)--(0.5,4)--(0.5, 5)--(1,5);
      \end{tikzpicture}
   }; 
\end{equation}
they do not commute with each other. 
Subsequently, the random average of the square of the norm can be expressed 
as the sum of all arrangement patterns 
of the non-commutative symbols $\mathcal{A}$ and $\mathcal{B}$: 
\begin{align}
   \overline{\braket{\psi_\beta|\psi_\beta}^2} = 
   \mathcal{A}\mathcal{A}\cdots\mathcal{A} + 
   \mathcal{B}\mathcal{A}\cdots\mathcal{A} + 
   \mathcal{A}\mathcal{B}\cdots\mathcal{A} + \cdots . 
\end{align}
Since $\mathcal{A}^N$ is identical to $\left(\overline{\braket{\psi_\beta|\psi_\beta}}\right)^2$, 
the variance of the norm is the sum of arrangements that include at least one $\mathcal{B}$: 
\begin{align}
   \mathrm{Var}\left(\braket{\psi_\beta|\psi_\beta}\right) = 
   \mathcal{B}\mathcal{A}\cdots\mathcal{A} + 
   \mathcal{A}\mathcal{B}\cdots\mathcal{A} + \cdots + 
   \mathcal{B}\mathcal{B}\cdots\mathcal{B}. 
\end{align}
In order to extract the leading terms of $\chi$ from this summation, we examine the pairs of $\mathcal{A}$ and $\mathcal{B}$. 
The pairs $\mathcal{A}\mathcal{A}$ and $\mathcal{B}\mathcal{B}$ each form two loops 
that contribute to the value $\chi^2$. 
On the other hand, the pairs $\mathcal{A}\mathcal{B}$ and $\mathcal{B}\mathcal{A}$ produce a single loop representing $\chi$.
As a result, the leading terms have fewer ``domain walls'' for $\mathcal{A}$ and $\mathcal{B}$, 
and the most leading terms for $\chi$ include only one ``domain'' for $\mathcal{B}$. 
By summing up the terms with single domain $\mathcal{B}^l$, we derive
\begin{align}
   &\mathrm{Var}\left(\braket{\psi_\beta|\psi_\beta}\right) \notag \\
   &= \chi^N \sum_{l=1}^N (N-l+1) z(\beta)^{2(N-l)} z(2\beta)^l 
   + \mathcal{O}\left(\chi^{N-2}\right). 
\end{align}
To compute the NFPF, we normalize the variance of the norm as 
\begin{align}
   \delta z^2 
   &= \frac{1}{\chi^2} \sum_{l=1}^N (N-l+1) 
   \left\{\frac{z(2\beta)}{z(\beta)^2}\right\}^l
   + \mathcal{O}\left(\frac{1}{\chi^4}\right).
\end{align}
This equation represents a series expansion 
that can be mathematically evaluated and can be simplified 
when we consider $z(2\beta)/z(\beta)^2 = e^{-s_2(\beta)}$: 
\begin{align}
   \delta z^2 
   &= \frac{1}{\chi^2} \sum_{l=1}^N (N-l+1) e^{-ls_2(\beta)}
   + \mathcal{O}\left(\frac{1}{\chi^4}\right) \notag \\
   &= \frac{1}{\chi^2} \left\{\frac{N}{e^{s_2(\beta)}-1} - 
   \frac{1-e^{-Ns_2(\beta)}}{(e^{s_2(\beta)}-1)^2}\right\}
   + \mathcal{O}\left(\frac{1}{\chi^4}\right). 
   \label{eq:nfpf_formula1}
\end{align}

\subsection{Correction factor for interactions}
\label{subsec:classical_ising}
In the previous section, 
an exact formula for the NFPF of the noninteracting Hamiltonian was derived. 
Now, based on this, we derive the formula of the NFPF for general interacting Hamiltonians, 
which can be done by introducing a single correlation factor as a numerically 
determined parameter. 
\par
We first analytically investigate the NFPF for the 1D classical Ising model 
to generate an idea of correlation factors without bias or approximations. 
The Hamiltonian is given as
\begin{equation}
    H = - J \sum_{i=1}^{N-1} \sigma^z_i \sigma^z_{i+1}, 
\end{equation}
which is the $g=0$ limit of the transverse-field Ising model 
that will be introduced shortly. 
The partition function is represented by the transfer matrix as 
\begin{equation}
    Z(\beta) = \bm{v}^\top X^{N-1} \bm{v}, 
\end{equation}
where 
\begin{equation}
    X(\beta) = 
    \begin{bmatrix}
        e^{\beta J} & e^{-\beta J} \\
        e^{-\beta J} & e^{\beta J}
    \end{bmatrix}, \quad 
    \bm{v} = 
    \begin{bmatrix}
        1 \\ 1 
    \end{bmatrix}. 
\end{equation}
The free energy in the thermodynamic limit can be computed 
thorough the maximal eigenvalue of the transfer matrix: 
\begin{align}
    f(\beta) 
    &= - \frac{1}{\beta} \lim_{N\to\infty} \frac{1}{N} \log Z(\beta) \notag \\
    &= - \frac{1}{\beta} \log (2 \cosh \beta J). 
\end{align}
The thermal R\'{e}nyi-2 entropy is obtained as
\begin{align}
    s_2(\beta) 
    &= 2\beta \{ f(2\beta) - f(\beta) \} \notag \\
    &= \log \left( 1 + \frac{1}{\cosh(2\beta J)} \right).
\end{align}
\par
Next, we calculate a $\chi$-leading term that appeares in the NFPF. 
The leading term of the present interacting case can be discussed 
in the same manner as the one given for the noninteracting case in the previous section. 
First, we separate the whole system into $A$ and $B$; 
as subsystem $A$, the first $m$ sites and the last $n$ sites 
are taken, and for subsystem $B$, we have 
the center $l$ sites with $m+n+l=N$. 
Then, the leading term corresponding to this bipartition is represented as 
\begin{equation}
    \frac{1}{\chi^2} \frac{\Tr_B \left[ \left( \Tr_A e^{-\beta H} \right)^2 \right]}
    {\left( \Tr e^{-\beta H} \right)^2}. 
\end{equation}
By defining a tensor $Y$ as 
\begin{equation}
    _{s_1}Y(\beta)^{s_2}_{s_3} = \exp[\beta J s_1 (s_2 + s_3)], 
\end{equation}
the numerator can be expressed as 
\begin{align}
    &\Tr_B \left[ \left( \Tr_A e^{-\beta H} \right)^2 \right] \notag \\
    &= 
    \begin{matrix}
        \bm{v}^T X(\beta)^{m-1} \\
        \bm{v}^T X(\beta)^{m-1}
    \end{matrix}
    Y(\beta)^T X(2\beta)^{l-1} Y(\beta) 
    \begin{matrix}
        X(\beta)^{n-1} \bm{v} \\
        X(\beta)^{n-1} \bm{v}
    \end{matrix}. 
\end{align}
Consequently, the leading term is calculated as 
\begin{align}
    \frac{\Tr_B \left[ \left( \Tr_A e^{-\beta H} \right)^2 \right]}
    {\left( \Tr e^{-\beta H} \right)^2} 
    = \frac{e^{-(l-1)s_2(\beta)}}{2}. 
\end{align}
The NFPF of the classical Ising chain is obtained 
by replacing $e^{-ls_2(\beta)}$ 
with $e^{-(l-1)s_2(\beta)}/2$
in Eq.~\eqref{eq:nfpf_formula1}: 
\begin{align}
    &\delta z^2 
    = \frac{1}{\chi^2} \sum_{l=1}^N (N-l+1) \frac{e^{-(l-1)s_2(\beta)}}{2}
   + \mathcal{O}\left(\frac{1}{\chi^4}\right) \notag \\
   &= \frac{1}{\chi^2} \frac{e^{s_2(\beta)}}{d}
   \left\{\frac{N}{e^{s_2(\beta)}-1} - 
   \frac{1-e^{-Ns_2(\beta)}}{(e^{s_2(\beta)}-1)^2}\right\}
   + \mathcal{O}\left(\frac{1}{\chi^4}\right). 
   \label{eq:nfpf_classical}
\end{align}
Here, $d$ is local degrees of freedom; 
in the case of spin-half chains, $d=2$. 
\par
The two cases we obtained analytically without bias, Eqs.~\eqref{eq:nfpf_formula1} 
and \eqref{eq:nfpf_classical}, have most parts of their forms in common. 
It is then natural to introduce a real parameter $\alpha$ 
to bridge them as 
\begin{equation}
    \delta z^2 
   \simeq \frac{1}{\chi^2} 
   \left( \frac{e^{s_2(\beta)}}{d} \right)^\alpha
   \left\{\frac{N}{e^{s_2(\beta)}-1} - 
   \frac{1-e^{-Ns_2(\beta)}}{(e^{s_2(\beta)}-1)^2}\right\}. 
   \label{eq:nfpf_formula}
\end{equation}
We expect this form to be valid for other quantum many-body systems represented by TPQ-MPS. 
Indeed, this form is justified by the following numerical demonstrations 
in which we treat $\alpha$ as a fitting parameter. 
\begin{figure*}
    \centering
    \includegraphics[width=\hsize]{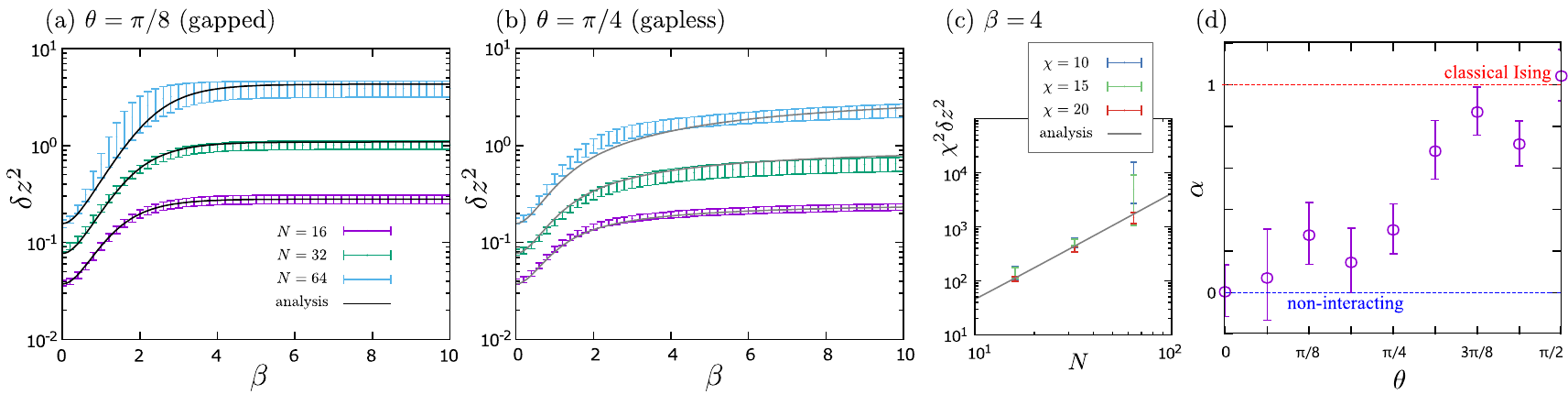}
    \caption{
    NFPF $\delta z^2$ obtained with the TPQ-MPS method 
    as a function of inverse temperature $\beta$ for the transverse-field Ising chain 
    at (a) $\theta = \pi /8$ (gapped paramagnetic phase) 
    and (b) $\theta = \pi /4$ (gapless critical point). 
    We use $N=16, 32, 64$ and $\chi = 20$ and take $M=500$ samples for each data point. 
    The error bars are calculated by a jackknife analysis. 
    The solid lines are the analytical results
    in Eq.~\eqref{eq:nfpf_formula} with fitting parameter 
    (a) $\alpha \simeq 0.28$ and (b) $\alpha \simeq 0.33$. 
    (c) The system size dependence of $\chi^2 \delta z^2$ 
    for $\theta = \pi /8$ and $\chi = 10, 15, 20$ at $\beta=4$. 
    (d) The fitting parameter $\alpha$ 
    for different model parameters $\theta$. 
    The error bars are calculated based on the associated errors in the NFPF.}
    Dashed lines show analytical conjectures for the 
    two limiting cases, $\theta =0$ and $1$, 
    which are the noninteracting and classical Ising models, respectively. 
    \label{fig:tfising}
\end{figure*}
\subsection{Crossover of sample complexity}
\label{subsec:crossover}
It is important to note that the NFPF in Eq.~\eqref{eq:nfpf_formula} 
exhibits different scaling forms in relation to the system size in the high- and low-temperature regions. 
When $N s_2(\beta) \gg 1$, we can discard the term $e^{-Ns_2(\beta)}$, 
and Eq.~\eqref{eq:nfpf_formula} becomes
\begin{equation}
    \delta z^2
    \simeq \frac{N}{\chi^2} 
    \left( \frac{e^{s_2(\beta)}}{d} \right)^\alpha
    \frac{1}{e^{s_2(\beta)}-1} 
    \label{eq:scaling_highT}
\end{equation}
in the large-$N$ limit. 
The NFPF scales linearly with the system size $N$
at high temperature. 
On the other hand, when $N s_2(\beta) \ll 1$, 
the exponential functions can be expanded as 
\begin{gather}
    e^{-Ns_2(\beta)} \simeq 1 - Ns_2(\beta) + \frac{N^2 s_2(\beta)^2}{2} \\
    e^{s_2(\beta)} \simeq 1 + s_2(\beta) + \frac{s_2(\beta)^2}{2}. 
\end{gather}
Therefore, Eq.~\eqref{eq:nfpf_formula} becomes
\begin{equation}
    \delta  z^2 \simeq
    \frac{N^2}{\chi^2} \frac{1}{2d^\alpha} 
    \label{eq:scaling_lowT}
\end{equation}
in the large-$N$ limit. 
The NFPF is proportional to the square of the system size $N$ 
at low temperature. 
\par
The crossover inverse temperature $\beta_c$ 
is characterized by $N s_2(\beta_c) \sim 1$. 
Because the thermal R\'{e}nyi-2 entropy scales as 
$s_2(\beta) \sim e^{-\beta \Delta E}$ in the low-temperature limit, 
where $\Delta E$ is the spectral gap, 
the crossover inverse temperature can be estimated as 
\begin{equation}
    \beta_c \sim \frac{1}{\Delta E} \log N. 
    \label{eq:betac}
\end{equation}
The crossover of the NFPF appears as a finite-size effect. 
\section{Numerical demonstrations}
In this section, 
we demonstrate the validity of Eq.~\eqref{eq:nfpf_formula} in two 
quantum many-body models. 
At each TEBD step, the bond dimension $\chi$ is determined so as to have 
the truncation error smaller than $10^{-8}$. 
We used the ITensor library for the MPS calculations \cite{itensor}. 

\subsection{Transverse-field Ising chain}
We first consider the transverse-field Ising chain, 
\begin{equation}
    H = - J \sum_{i=1}^{N-1} \sigma_i^z \sigma_{i+1}^z
    - g \sum_{i}^N \sigma_i^x, 
\end{equation}
where $\sigma^\alpha_i ~ (\alpha = x, y, z)$ is the Pauli operator at site $i$. 
We parametrize the coupling interaction $J$ 
and the transverse field $g$ as $J = \sin \theta$ and $g = \cos \theta$ 
with a single parameter $\theta$ that varies 
in the range $[0, \pi/2]$. 
It is analytically solved in the thermodynamic limit
\cite{pfeuty1970}. 
When $J=0$ and $g>0$, the Hamiltonian is represented as a sum of single-site operators 
treated in Sec.\ref{subsec:non-interacting}, and when $J>0$ and $g=0$, 
we find the 1D classical Ising model we considered in Sec.\ref{subsec:classical_ising}. 
The noninteracting and classical Ising cases correspond 
to $\theta = 0$ and $\theta = \pi/2$, respectively. 
When $0 \le \theta < \pi /4$, we have a gapped paramagnetic phase due to large $g$; 
for $\pi /4 < \theta \le \pi /2$, the system is in a gapped ferromagnetic Ising phase, 
and the gap closes at $\theta=\pi/4$. 
Whether the system is gapped or gapless is related to the nature of the NFPF. 
\par
Figures~\ref{fig:tfising}(a) and \ref{fig:tfising}(b) show 
the temperature dependence of the NFPF $\delta z^2$ 
for $\theta = \pi/8$ and $\pi/4$, respectively. 
The solid lines represent the formula in Eq.~\eqref{eq:nfpf_formula}. 
The thermal R\'{e}nyi-2 entropy $s_2(\beta)$ of the transverse-field Ising chain 
can be analytically calculated from an exact result, 
and using its value, we can fit the mean value of the numerically obtained NFPF data 
by a single parameter $\alpha$. 
Here, the error bars are calculated based on the associated errors in the NFPF.
In the gapped paramagnetic case in Fig.~\ref{fig:tfising}(a), 
using $\alpha\simeq 0.28$, which is common to three choices of $N$, 
the formula and the data agree well throughout the whole temperature range. 
All NFPFs increase at high temperature and reach a plateau at low temperature. 
This behavior indicates the fact that 
low- and high-temperature regions follow different scaling laws, 
as expected from Eqs.~\eqref{eq:scaling_highT} and \eqref{eq:scaling_lowT}. 
As discussed in Sec.\ref{subsec:crossover}, 
the crossover temperature $\beta_c$ in Eq.~\eqref{eq:betac} 
grows logarithmically with $N$, consistent with Fig.~\ref{fig:tfising}(a). 
\par
Figure.~\ref{fig:tfising}(b), the gapless case, 
also shows good agreement with the analytical result, 
but there are no plateau regions in contrast to the above example. 
This is because $\beta_c$ diverges in gapless systems with $\Delta E=0$. 
\par
Figure.~\ref{fig:tfising}(c) shows 
the system size dependence of $\chi^2 \delta z^2$. 
Because $N$ appears in the numerators of the scaling formulas 
\eqref{eq:scaling_highT} and \eqref{eq:scaling_lowT},  Eq.~\eqref{eq:nfpf_formula} is considered to be 
an expansion for large bond dimensions and small system size. 
\par
In Fig.~\ref{fig:tfising}(d), we show the evolution 
of the fitting parameter $\alpha$ in Eq.~\eqref{eq:nfpf_formula} 
as a function of $\theta$. 
Here, $\alpha$ is calculated for $N=16$ and $\chi=20$. 
In the noninteracting and classical Ising model, 
numerical results and analytical estimation agree well. 
The fitting parameter $\alpha$ interpolates the two limits 
within the error bars, where we find a nonmonotonous change at $\theta\sim \pi/4$, 
at which the ground state becomes gapless. 
We may thus speculate that the value of $\alpha$ is determined by the nature of the phase, 
while the form of Eq.~\eqref{eq:nfpf_formula} is safely kept 
throughout the whole temperature range. 

\subsection{Heisenberg chain}
\begin{figure}
    \centering
    \includegraphics[width=0.7\hsize]{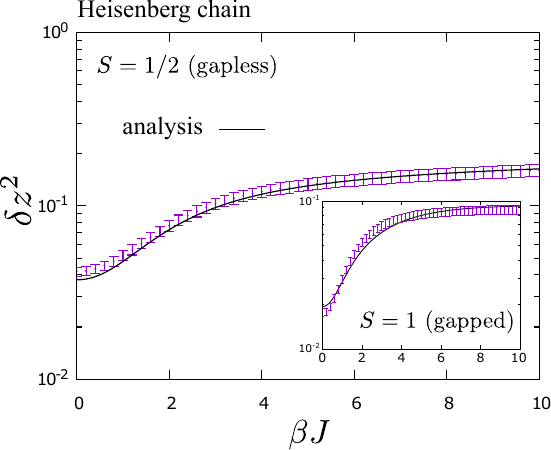}
    \caption{
    NFPF $\delta z^2$ of $S=1/2$ and $S=1$ (inset) Heisenberg chains 
    for $N=16, \chi=20$.
    The solid lines represent the analytical result in Eq.~\eqref{eq:nfpf_formula} 
    for fitting parameters $\alpha \simeq 0.66$ and $\alpha \simeq 1.25$ (inset). 
    }
    \label{fig:heisenberg}
\end{figure}
We next apply the Heisenberg chain defined as 
\begin{equation}
    H = J \sum_{i=1}^{N-1} \bm{S}_i \cdot \bm{S}_{i+1}, 
\end{equation}
where $S^\alpha_i ~ (\alpha =x, y, z)$ 
is the spin operator at site $i$. 
The $S=1/2$ Heisenberg model is exactly solvable \cite{bethe1931}, 
and the ground state is gapless. 
The $S=1$ case has a Haldane gap of $\Delta E \simeq 0.4105$ 
\cite{haldane1981, haldane1983, white1993-2}. 
We treat both cases to compare gapless and gapped systems. 
Figure~\ref{fig:heisenberg} shows the results for $N=16$ and $\chi=20$, 
both of which are consistent with the analytical formula. 
The thermal R\'{e}nyi-2 entropy used in the fitting function 
is evaluated by the TPQ-MPS method at $N=100$. 
We can find a plateau at low temperature in the inset of Fig.~\ref{fig:heisenberg}, 
which indicates the crossover of the system size scaling. 
The analytical formula in Eq.~\eqref{eq:nfpf_formula} 
is thus expected to hold. 

\section{Summary and discussion}
As one of the measures of difficulty in computing finite-temperature quantum states, 
we investigated the normalized fluctuation of the partition function (NFPF). 
This quantity is well-defined and measurable for a general random sampling method that 
generates a thermal state using the imaginary time evolution from the high-temperature random states. 
The NFPF is proportional to the required number of samples, 
and reflects the sample complexity of the calculation. 
Focusing on the matrix-product-state-based (MPS-based) random sampling method,  
we analytically derived an exact form of the NFPF in Eq.~\eqref{eq:nfpf_formula} 
and demonstrated its validity numerically for several representative models. 
Our formula shows a crossover behavior with varying temperature; 
at high temperatures, the NFPF scales linearly with the system size, 
while at low temperatures, it is proportional to the square of the system size.  
This result gives a practical and quantitative demonstration of the computational complexity 
of 1D systems shown in Fig.~\ref{fig:complexity}(a). 
\par
The computational cost of the MPS-based sampling relies on two quantities, 
the required number of samples and the memory cost per sample. 
Our result and the previous conformal field theory (CFT) calculations indicate, respectively, 
that both the former and the latter scale polynomially with the system size at any finite temperatures. 
Because the ground state is known to be QMA-complete, a question arises: 
How can such computationally feasible finite-temperature states bounded by polynomials 
continue to the zero-temperature limit? 
As we discussed in the Introduction, 
QMA-completeness may rely on several nontypical states in 1D models 
that deviate from CFT predictions at low temperatures 
such as the Motzkin chain \cite{bravyi2012, movassagh2016}. 
In these models, the bond dimension of the thermal state diverges as the temperature approaches zero. 
However, explicitly presenting such models as evidence would imply proving that P$\neq$PSPACE, 
a claim we cannot substantiate. 
\par
So far, we have considered the 1D models with nearest-neighbor interactions. 
However, we previously applied the TPQ-MPS method to the 2D Kitaev honeycomb model \cite{gohlke2023}, 
showing that the method works fairly well, reproducing two peaks of the specific heat 
that were previously established based on the specific Majorana description of the model. 
There, the MPS is constructed along the 1D path wrapping the cylinder in a spiral construction. 
Because such MPS construction converts the nearest-neighbor interactions of the original model 
to the longer-range interactions, this suggests that the present scaling relationship 
may be extended to wider classes. 
Typically, the long-range interactions obey $1/r^{\Tilde{\alpha}}$ with distance $r$ 
between two particles or spins. 
Increasing $\Tilde{\alpha}$ smoothly interpolates to the short-range interactions. 
If $\Tilde{\alpha} >2$, 
the ground state maintains an entanglement area law \cite{gong2017, kuwahara2020-2}, 
and because of the robust volume law at high temperatures, 
there should be a crossover between the two regimes.  
In such a case, we may expect our formula \eqref{eq:nfpf_formula} to remain applicable, 
and a crossover in the NFPF scaling law that detects 
the difference between the sampling from an ensemble and the sampling from a pure ground state may exist. 
Such long-range interacting systems are realized in experiments with trapped ions, cold atoms, and Rydberg atoms
\cite{defenu2023}.
\par
Finally, let us discuss the possibility of straightforwardly extending 
our analysis in Sec.\ref{subsec:non-interacting} to higher-dimensional systems. 
There, the $\mathcal{B}$ domain can no longer exist in the leading terms 
in the same manner because the classical Ising model requires finite energy to form domains in higher dimensions. 
Consequently, the NFPF becomes proportional to the system size, 
and the crossover observed in 1D no longer exists. 
However, it is questionable whether this extension has practical meaning. 
The entanglement in the ground state of higher dimensions is extremely complex compared to that in one dimension, 
and it remains a topic of ongoing debate; 
a similar situation can be expected in thermal states. 
If this is the case, our calculation in Sec.\ref{subsec:non-interacting}, 
which assumes a certain type of cluster property, likely fails. 
As the high-dimensional and low-temperature regime is associated 
with the unsolved quantum PCP conjecture in Fig.~\ref{fig:complexity}(a), 
it warrants careful discussion. 
\par
In this paper, we have focused on the random sampling methods using classical computers, 
but the conversion of the related classical algorithms to quantum algorithms is now ongoing 
\cite{motta2019, sun2021, seki2022, coopmans2023, goto2023, mizukami2023, davoudi2023, joao2023, pedersen2023}. 
Because they are demonstrated virtually on classical computers or 
on real quantum computers with very small sizes, so far it has been difficult to evaluate their effectiveness. 
In reality, in near-term quantum computers, the initial random states are prepared on random circuits 
of shallow depth, and there, 
the NFPF or the number of samples is expected to follow the same scaling as our results. 
\par
We finally note that our calculations in Sec.\ref{subsec:non-interacting} are similar to 
those treating the R\'{e}nyi-2 entropy of an evaporating black hole in Ref.~[\onlinecite{penington2022}], 
but the physical implications of these relationships are not yet well understood. 
Following the approach in Ref.~\cite{penington2022}, 
we can compute cumulants of any order for the norm of thermal states in MPS-based random sampling.
This analysis pertains to a more detailed efficiency of the method. 

\begin{acknowledgments}
We thank Kouichi Okunishi and Hosho Katsura 
for fruitful discussions. 
A. I. was supported by a Grant-in-Aid for JSPS Research Fellow (Grant No. 21J21992).
This work was supported by a Grant-in-Aid for Transformative Research Areas 
``The Natural Laws of Extreme Universe---A New Paradigm for Spacetime and Matter from Quantum Information" (Grant No. 21H05191)
and other JSPS KAKENHI Grants (Grant No. 21K03440), Japan. 
\end{acknowledgments}

\nocite{apsrev42Control}
\bibliographystyle{apsrev4-2}
\bibliography{myref}

\end{document}